\documentclass{aa}
\usepackage{graphics}

\begin{document}

   \title{Long Period Variables detected by ISO in the SMC}
   
   \author{M.-R.L. Cioni\inst{1}
        \and 
          J.A.D.L. Blommaert\inst{2}
        \and
          M.A.T. Groenewegen\inst{2}
        \and
          H.J. Habing\inst{3}
        \and
          J. Hron\inst{4}
        \and
          F. Kerschbaum\inst{4}
        \and
          C. Loup\inst{5}
        \and
          A. Omont\inst{5}
	\and
	  J.Th. van Loon\inst{9}
        \and
          P.A. Whitelock\inst{7}
        \and
          A.A. Zijlstra\inst{8}
        }
 
   \offprints{mrcioni@eso.org}

   \institute{European Southern Observatory,    
              Karl--Schwarzschild--Str.~2, 
              D--85748 Garching bei M\"{u}nchen, Germany
        \and
              Institute of Astronomy, Katholieke Universiteit Leuven, 
              Celestijnenlaan 200B, B3001, Leuven, Belgium 
        \and
              Sterrewacht Leiden, Niels Bohrweg 2,
              2333 RA Leiden, The Netherlands
        \and
              Institut f\"{u}r Astronomie der Universit\"{a}t Wien,
              T\"{u}rkenschanzstrasse 17, 1180 Wien, Austria
        \and
              Institute d'Astrophysique de Paris, CNRS, 
              98bis Bld.~Arago. 75014 Paris, France
        \and
	      Astrophysics Group, School of Chemistry and Physics,
	      Keele University, Staffordshire ST5 5BG, United Kingdom
        \and
              South African Astronomical Observatory, P.O.Box 9, 
              7935 Observatory, Republic of South Africa
        \and
              Department of Physics, UMIST, P.O. Box 88,
               Manchester M60 1QD, United Kingdom
             }

   \date{Received .../ Accepted ...}

   \titlerunning{ISO LPVs in the SMC}
  
   \authorrunning{M.-R. Cioni et al.}
   
   \abstract{This article presents the study of the light--curves 
     extracted from  the MACHO database of a sample of stars observed 
     by the Infrared Space Observatory in the 
     Small Magellanic  Cloud. These  stars belong to
     the   ISO--Mini-Survey  catalogue   of   the  Magellanic   Clouds
     (ISO--MCMS, Loup  et al. {\it in preparation}). Most of  them are  in the
     asymptotic giant branch (AGB) and supergiant phases. The dominant
     period and amplitude of pulsation have been derived and the stars
     have  been   classified  as  Mira   or  Semi--Regular  pulsators.
     Furthermore, the  cross--identification with near--infrared DENIS
     and 2MASS  magnitudes available within the  ISO--MCMS allowed us:
     (i)  to investigate  the  properties of  these  stars in  the
     combined  near--  and  mid--infrared colour--magnitude  diagrams,
     (ii)  to  derive  the  bolometric magnitude  by  integrating  the
     spectral   energy  distribution  and   (iii)  to estimate  the
     mass--loss rate. The stars  have been divided into carbon-- (C--)
     and  oxygen--rich  (O--rich) using  the  ($J-K_S$, $K_S$)  
  colour--magnitude  diagram and their  period and  amplitude distributions
     have  been compared.  C--rich  AGB stars have a sharp peak in their 
     period distribution at 
     about  $250$ days  and have  on average  a larger  amplitude than
     O--rich AGB stars.  This effect, not previously detected from the
     study of similar  stars in the Large Magellanic  Cloud and in the
     Baade's window,  might be closely  related to the  metallicity of
     the environment in which the stars have formed.}

\maketitle

\section{Introduction}
The stellar content of the Small Magellanic Cloud (SMC) can be studied
in  great detail  because of  the  low galactic  extinction and a known
distance   m$-$M$=18.99\pm0.03$(formal)$\pm0.08$(systematic),  Cioni  et
al.  (\cite{cionitip}).  However,  it  is  well  known,  but  not  yet
quantified, that the SMC is extended along the line of sight much more
than its companion galaxy,  the Large Magellanic Cloud (LMC). Together
with our  own Galaxy, they  constitute an in\-ter\-act\-ing system  (e.g.  
Gardiner \& Noguchi \cite{garden}, Moore \& Davis \cite{moore}, 
Heller \& Rohlfs \cite{heller})
and a  metallicity scale environment in  which the Galaxy  is the most
metal  rich and  the SMC  the  most metal  poor of  the three. Fields 
covering  almost  the  whole  of  the  two  Magellanic  Clouds  and  a
considerable  number of  fields  in the  Galaxy  have been  repeatedly
observed by the micro--lensing projects:  OGLE, MACHO and EROS. This is
providing a  wealth of  data to identify  and study variable  stars in
every  evolutionary phase.   Partial or  all  sky survey  data in  the
near--infrared (IR) provide  essential counterparts for the specific
a\-na\-ly\-sis of the reddest objects: red (RGB) and asymptotic giant branch
(AGB) stars. Most AGB stars are pulsating with an amplitude of up to a
few  magnitudes in the  optical wave  bands and  somewhat less  in the
near--infrared bands. Long  Period Variables (LPVs) have a period of 
at least $100$ or $150$ days and longer. Many  are also
multi--periodic objects.  More than  one dependence between period and
luminosity has been  detected in the LMC (Wood  et al. \cite{woodiau},
Cioni  et al. \cite{cionivar},  Noda et  al. \cite{moa},  Lebzelter et
al. \cite{agape}).  Each period--luminosity relation (PLR) is probably
associated with a different mode of stellar pulsation. Because most of
these stars  are probably experiencing  the thermal pulsing  AGB phase
(TP--AGB) they are losing mass at an average rate of 
$10^{-6}    $M$_{\sun}$yr$^{-1}$   and    they   have    highly   extended
atmospheres.  The pulsation as  the driving  mechanism of  the stellar
wind is a  fundamental element in the study of  the mass--loss rate of
AGB stars. They are surrounded  by an initially optically thin, and later 
optically thick,
circumstellar envelope  and thus sometimes become detectable  only in the
infrared wave bands.   Many obscured AGB stars were  discovered by the
IRAS  satellite. More recently the Midcourse Space Experiment (MSX - Price 
et al. \cite{msx}) observed selected high density regions in the Magellanic 
Clouds. Improved source position were derived. Egan et al. (\cite{egan})  
present the cross-identification between MSX and 2MASS data in the LMC. 
However, the  Infrared Space  
Observatory (ISO) obtained more  sensitive data in 
some  fields in  the Magellanic Clouds  and the  Galaxy that 
increased the  number of obscured AGB stars  and allowed  us to  learn 
in more  detail about  their dust properties.

In order to  study the evolved AGB population it  is thus important to
utilize observations that cover the spectral range from the optical
($I$  band)  to the  mid--IR.  It  is  then possible  to  derive
fundamental quantities  such as the bolometric magnitude  and the 
mass--loss rate  of the stars. In  addition, from the  period, the amplitude
and the  regularity of the  variation, it is possible  to cha\-rac\-te\-ri\-se
the stars  as Mira  or Semi--Regular (SR)  pulsators and  to indicate,
from the PLR, the mass of  the stellar progenitor. In this article the
properties of  a sample  of AGB  stars in the  SMC extracted  from the
ISO--Mini-Survey catalogue  of the Magellanic  Clouds (ISO--MCMS, Loup
et al. {\it in preparation}),  which already contains the cross--identification
with the near--IR  catalogues DENIS and 2MASS, are  discussed. This is done
by additionally  analysing the light--curves extracted  from the MACHO
database.    A  similar   work  was performed  in   the  LMC bar--west field   
(Cioni  et al. \cite{cionivar}) and will be further extended to other fields were
ISO measurements are  now available. A detailed comparison  of more or
less obscured  AGB va\-ria\-bles  between the two  Clouds and  the Baade's
window NGC6522 will  be given in another paper (Glass,
Schultheis \& Cioni, {\it in preparation}).

The  data and  the  cross--identification with  the light--curves  are
described  in Sect.~2.   In Sect.~3  the  analysis of  the period  and
amplitude  of  variability and  the  determination  of the  bolometric
magnitude is  presented.  A discussion of the  different type
and variability class of the sources in the PLRs and in the near-- and
mid--IR colour--magnitude diagrams  is given in Sect.~4.  Sect.~5
concludes the article. Appendix A discusses the re--analysis of the LPVs
studied  by  Feast  et  al.  (\cite{feastpl})  and  B  compares  the
bolometric  correction obtained  in this  work with  the  relations by
Alvarez et al. (\cite{alva}) and Montegriffo et al. (\cite{monte}).

\section{Data}


The  sample  of data  in  the  SMC, analysed  here,  is  based on  the
cross--identification between the ISO--MCMS (Loup et al. {\it in preparation}),
 and the  MACHO light--curves publicly
available\footnote{http://wwwmacho.mcmaster.ca}.

The ISO  mini--survey in the  direction of the  SMC co\-vers an  area of
$0.28$ square  degrees. Observations were performed  with the infrared
camera ISOCAM on board the ISO  satellite in the LW2 ($6.61 \mu$m) and
LW10  ($11.18 \mu$m)  wavebands.  The catalogue  lists $1333$  sources
brighter  than  about   $12.4$  and  $11.3$  mag  in   LW2  and  LW10,
respectively.   The coordinates  of the  ISO sources  are  better than
$4$\arcsec~with  a faint  tail up to  $6$\arcsec.  The  catalogue also
lists  the cross--identification  with  the near--infrared  catalogues
2MASS               (2nd              Incremental              Release
PSC\footnote{http://www.ipac.caltech.edu/2mass/})  and the DENIS 
(Epchtein et al. \cite{nicolas}) catalogue 
towards the Magellanic Clouds (DCMC -- Cioni et
al. \cite{dcmc}), and a few  smaller optical catalogues of carbon stars,
HII regions, PNe and emission--line stars. According to the association
likelihood criteria defined by Loup et al. ({\it in preparation}) $48\%$ 
of the
sources have  no, or an unlikely, association  with DENIS/2MASS; $13\%$
have a  questionable association; $25\%$ have  a confident association
and the remaining  $14\%$ are ``known'' sources with  a DENIS/2MASS and
optical counterpart. It is perhaps surprising that there are many 
ISO sources without a near--infrared counterpart. However, most of these 
sources are among the faintest ISO detections in both ISO wavebands. 
The histogram of their magnitudes and colour peak at: $LW2=12$, $LW10=11$ 
and $(LW2-LW10)=1.5$. Moreover, the ISO--MCMS includes sources well 
detected in one band only. Among sources detected only in two bands about 
$86\%$ have a near--IR counterpart. The ISO photometric magnitudes are 
part of the final version of the catalogue released to the Co--Is and 
associates on March 2002. 

The MACHO  (Alcock et al., \cite{alcock})  light--curves were obtained
 from  repeated measurements in  two, blue  and red,  broad pass--bands
 from  the 50  inch telescope  at the  Mount Stromlo  Observatory in
 Australia.  Observations  of fields  covering most of  the Magellanic
 Clouds  were  performed  from $1992$  to
 $2000$. The light--curves are easily accessible via the web interface
 giving in input the right ascension (RA) and the declination (DEC) of
 each source.

\begin{table*}
\caption{Extract of the table of MACHO counterparts of ISO--MCMS SMC sources}
\label{machoid}
\[
\begin{array}{ccccrlc}
\hline
\noalign{\smallskip}
\mathrm{RA} & \mathrm{DEC} & \mathrm{field} & \mathrm{tile} & \mathrm{seq.} &
\mathrm{dist.} & \mathrm{ISO-MCMS}\\ 
0:48:44.26 & -73:21:22.16 & 212 & 15903 & 16  & 0.22 & J004844.4-732119 \\
0:48:44.67 & -73:17:55.01 & 212 & 15904 & 27  & 1.02 & J004844.5-731754 \\
0:48:44.29 & -73:20:19.91 & 212 & 15904 & 5   & 2.36 & J004844.7-732018 \\
0:48:49.26 & -73:16:25.11 & 212 & 15905 & 525 & 1.12 & J004849.4-731627 \\
0:48:49.91 & -73:20:04.07 & 212 & 15904 & 6   & 1.08 & J004849.8-732002 \\
0:48:51.10 & -73:21:40.58 & 212 & 15903 & 11  & 0.71 & J004850.7-732138 \\
0:48:51.75 & -73:22:39.51 & 212 & 15903 & 1   & 0.55 & J004851.8-732239 \\
0:48:54.13 & -73:15:58.40 & 212 & 15905 & 523 & 1.43 & J004854.0-731557 \\
0:48:59.26 & -73:11:53.24 & 212 & 15906 & 2   & 1.26 & J004859.7-731155 \\
0:49:0.221 & -73:22:23.62 & 212 & 15903 & 9   & 0.71 & J004900.4-732224 \\
\noalign{\smallskip}
\hline
\end{array}
\]
\end{table*}

\begin{figure}
\resizebox{\hsize}{!}{\includegraphics{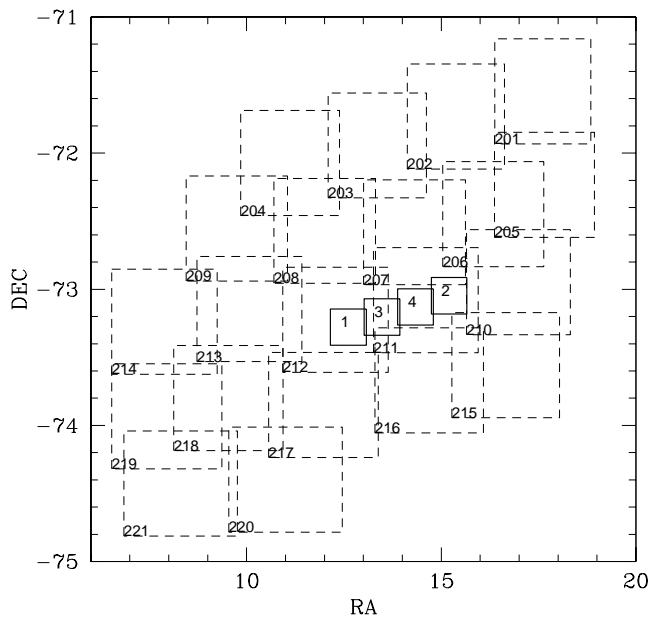}}
\caption{Location  of the  fields observed  by ISO  (continuous  line) and
MACHO (dashed line) in the SMC. The ISO fields lie in the bar of the galaxy 
and are each $16\arcmin\times16\arcmin$ in size. Both right ascension (RA) and 
declination (DEC) are in degrees.}
\label{regions}
\end{figure}

Fig. \ref{regions}  shows the regions observed by  ISO superimposed on
the MACHO  fields.  The area covered  by ISO is  entirely contained in
 MACHO fields  number  $211$ and  $212$;  a few  sources are  also
present in  field number $207$.  These MACHO fields were  observed for
the whole  period of eight years  while the observations  of fields in
the  external  parts  of the  SMC  cover  just  five years.  The  OGLE
light--curves cover a period  of about five years (observations started
in 1997). Because of the nature of LPVs, sources with periods from $100$ 
to  $1000$  days,   the  MACHO  light--curve  database  was
preferred.

\subsection{Match between ISO--MCMS and MACHO}

\begin{figure}
\resizebox{\hsize}{!}{\includegraphics{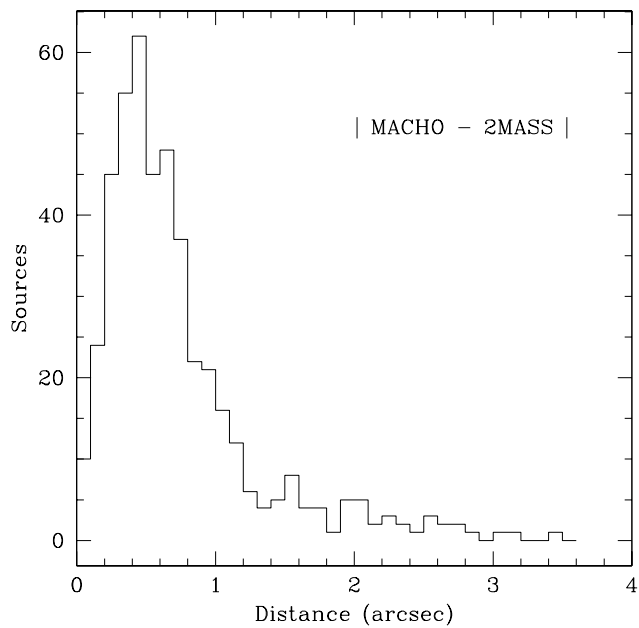}}
\caption{Histogram of the absolute distance between the 2MASS counterpart 
of an ISO--MCMS source and the corresponding MACHO source.}
\label{coord}
\end{figure}

Each  ISO source  with  a confident  DENIS/2MASS  association has  been
searched for a light--curve in the MACHO database using, as input, the
2MASS  coordinates.  Variables  were confidently  identified  within a
radius of $1$\arcmin.  Fig. \ref{coord} shows that most of the sources
were actually associated  within $1$\arcsec.  In a very  few cases the
MACHO counterpart, a  source with a typical LPV  light--curve, was not
the  closest to  the 2MASS  source --  but still  within no  more than
$4$\arcsec. We
consider  a light--curve  to be  that of  a LPV  when a  more  or less
periodic behaviour is detected by eye.
It is expected that most of the sources detected in the near-- and mid--IR 
are associated to AGB stars. This is a consequence of the mass--loss
process occurring in AGB stars that builds up a circumstellar envelope
observable in the mid--infrared.

Among the $502$ sources  with a confident DENIS/2MASS counterpart, $458$
have been  found to have  a MACHO light--curve.   Table \ref{machoid},
electronically  available, lists  the MACHO  coordinates at  the epoch
J2000  ({\it Column 1  \& 2}),  the MACHO  identifier field,  tile and
number ({\it  Column 3, 4  \& 5}), the  distance in arcsec  between the
2MASS  and  the  MACHO  source  ({\it  Column  6})  and  the  ISO--MCMS
identifier  ({\it Column 7}).  The first  $10$ lines  are given  as an
example (Table \ref{machoid}).

\begin{table}
\caption{Table of ISO--MCMS sources with a confident DENIS/2MASS counterpart but without an associated MACHO light--curve.}
\label{nocurves}
\[
\begin{array}{cl}
\hline
\noalign{\smallskip}
\mathrm{ISO-MCMS} & \mathrm{MACHO\,counterpart\,?}\\
J004850.9-731402 & d > 17\arcsec \\
J004902.7-732141 & d > 5\arcsec, \mathrm{only\, 3\, data\, points}\\ 
J004907.2-731335 & d > 12\arcsec, \mathrm{too\, few\, data\, points}\\
J004907.8-730920 & d > 14\arcsec, \mathrm{too\, few\, data\, points}\\
J004908.2-731415 & d > 14\arcsec \\
J004909.5-731133 & d > 7\arcsec \\
J004910.5-731859 & d > 3\arcsec, \mathrm{too\, few\, data\, points}\\
J004934.4-731409 & d > 9\arcsec \\
J004940.8-731352 & d > 15\arcsec, \mathrm{too\, few\, data\, points}\\
J004940.9-731412 & d > 8\arcsec \\
J004959.0-731058 & d > 7\arcsec \\
J005001.7-731349 & d > 9\arcsec \\
J005034.2-731354 & d > 15\arcsec \\                     
J005108.1-731343 & d > 5\arcsec \\
J005117.5-731659 & d > 4\arcsec, \mathrm{too\, few\, data\, points}\\
J005121.1-731416 & d > 4\arcsec \\
J005129.7-731044 & d > 8\arcsec \\
J005131.4-732008 & d > 5\arcsec, \mathrm{too\, few\, data\, points}\\
J005140.2-731333 & \\
J005202.5-731339 & d > 6\arcsec \\
J005207.0-731407 & d > 7\arcsec \\
J005211.5-731351 & d > 16\arcsec \\
J005217.8-731327 & d > 4\arcsec \\
J005218.9-730921 & d > 4\arcsec, \mathrm{too\, few\, data\, points}\\
J005230.8-730259 & d > 8\arcsec, \mathrm{too\, few\, data\, points}\\
J005239.9-730912 & \\
J005245.7-730425 & d > 6\arcsec \\
J005302.9-730746 & d > 6\arcsec \\
J005306.4-730633 & d > 10\arcsec \\
J005309.0-730404 & d > 9\arcsec \\
J005309.9-731324 & d > 3\arcsec, \mathrm{too\, few\, data\, points}\\
J005310.3-730722 & d > 3\arcsec, \mathrm{too\, few\, data\, points}\\
J005405.7-730459 & d > 3\arcsec, \mathrm{too\, few\, data\, points}\\
J005417.7-730504 & d > 3\arcsec, \mathrm{poor\, sampling}\\
J005418.1-730523 & d > 7\arcsec \\
J005518.6-731420 & d > 2\arcsec, \mathrm{too\, few\, data\, points}\\
J005549.0-730257 & d > 4\arcsec, \mathrm{too\, few\, data\, points}\\
J005646.4-730519 & d > 12\arcsec \\
J005648.0-730404 & d > 5\arcsec \\
J005650.1-731040 & d > 2\arcsec, \mathrm{too\, few\, data\, points}\\
J005650.9-730420 & d > 7\arcsec \\          
J005749.0-730522 & d > 11\arcsec, \mathrm{too\, few\, data\, points}\\
J005822.0-730734 & d > 4\arcsec \\           
J005833.1-730528 & d > 16\arcsec \\
\noalign{\smallskip}
\hline
\end{array}
\]
\end{table}

The sources listed  in Table \ref{nocurves}  are those for which  we could
not find a  convincing MACHO counterpart. The ISO--MCMS  name is given
in {\it Column 1}, the distance  in arcsec to the first possible MACHO
counterpart is given in {\it Column 2} and in some cases the reason of
the missing  cross--identification is written in {\it  Column 3}.  Too
few data points means that there are not enough points to characterise
the shape  of the  light--curve. This would  be useful to  recognize a
MACHO counterpart within the ISO astrometric precision.

\section{Period analysis}
The  Schwarzenberg--Czerny method  (\cite{periods}) has  been  used to
detect  periodicities in  the light--curves  of the  selected sources,
independently for  the measurements in  the blue and red  bandwidth of
MACHO.  The algorithm was executed thirty times  searching for periods 
in a  range from
$10$ to  $100$, $200$, ..., $3000$ days with a step of  $0.005$ days.  The
result of the  program is the period of the variation  and a few other
parameters  among  which the  quality  of  the  variability ``q''.  We
identified as the  main period the value corresponding  to the highest
value of  ``q'' among the thirty different determinations. 
We noticed  that at the  first passage through the  periodic detection
algorithm  the  dominant  period  did  not  give,  in  some  cases,  a
sufficiently defined phase--curve (distribution of the measurements 
folded on a given period).  By restricting  or enlarging  the  range of
searching periods it  was always possible to optimize  the phase curve
to a  certain period.  Therefore we decided  to execute  the algorithm
more times  by changing the period searching  window.  The reliability
of the assigned period was then inspected by looking at the phase curve
of  each source. Sources  of regular variability  and with a 
large  amplitude of variability
have  a clean  phase  plot with  a  large S/N  ratio  that allows  the
variation of the position of  the maximum or minimum brightness and/or
the presence of bumps to be distinguished (sometimes identified as the
secondary  period). In  other cases  the S/N  was not  good  enough to
detect these secondary variations but the phase--curve was clear enough
to confirm the main period.
Then  we introduced
the  following criteria  to  flag the  sources  in each band. 
Flag$=0$  for sources that  do not have  data points in  a given
MACHO band.  Flag$=5$ for  sources with $q\leq 15$ or  a period
that coincides with the extremes  of the searching window in all $30$
cases (i.e. sources with a period of $10$ or $100$, $200$,... $2900$,
$3000$   days).  Their   light  curve   is  judged   to  be   of  poor
quality. Flag$=1$ and  Flag$=2$ for sources with $q>15$  and for which
the assigned period (P) differs not more than $10$ days  from the mean of
all the measurements  (among the thirty possible) with  $q>15$ in the blue
and red band width, respectively.  With these criteria we are confident
that a  single periodicity is the  do\-mi\-nant source of  variation in the
light--curve.  Flag$=6$ for sources  with $q>15$ and a  difference from
the same mean  larger than $10$  days.  These  sources are
likely to have more than one significant periodicity.  In this work we
are not  aiming at  a detail analysis  of the secondary  periods which
will  be discussed,  together with  the  comparison of  LPVs with an ISO
counterpart between the  Magellanic Clouds and the Galaxy  , in Glass,
Schultheis \&  Cioni ({\it in preparation}).   Aliases were identified
from the  diagram $Log(P)$ versus  $K_S$ as those periods creating clear
vertical patterns. These correspond to  periods equal to $29$ days and
in the  range from $340$ to  $390$ days, included.  These sources were
assigned  Flag$=9$. Examples of  MACHO light--curves  for LPVs  can be
found in Wood et al.~(\cite{woodiau}).  Table \ref{periods} summarises
the detected periodicities. It lists the description of the Flag ({\it
Column 1}), the  value of the flag and the number of
sources corresponding to a given flag in the red ({\it Column 2 \& 3}) 
and blue bandwidths ({\it Column 4 \& 5}).

\begin{table}
\caption{Summary of detected periodicities}
\label{periods}
\[
\begin{array}{lcrcr}
\hline
\noalign{\smallskip}
\mathrm{Description} & \mathrm{Flag} & \mathrm{N} & \mathrm{Flag} & \mathrm{N}\\
 & \multicolumn{2}{c}{\mathrm{RED}} & \multicolumn{2}{c}{\mathrm{BLUE}} \\
\mathrm{No\,curve}       & 0 &   8 & 0 &   9 \\
\mathrm{Good}            & 1 &  96 & 2 &  91 \\
\mathrm{Poor}            & 5 & 103 & 5 & 208 \\
\mathrm{Multi-Periodic}  & 6 &  91 & 6 &  91 \\
\mathrm{Alias}           & 9 & 160 & 9 &  59 \\
\noalign{\smallskip}
\hline
\end{array}
\]
\end{table}

Table  \ref{general}  provides   the  magnitudes  from  the  different
databases,  the  bolometric  magnitude  (see  next  Section)  and  the
parameters that describe the dominant  pulsation of all the sources in
the sample with an extracted light--curve ($458$).  This table is also
made electronically available and the first $10$ lines are given as an
example  (Table  \ref{general}).  The  content  of the  table  is  the
following: ISO--MCMS  name ({\it  Column 1}), $IJK_S$  DCMC magnitudes
({\it Columns 2,  3 \& 4}), $JHK_S$ 2MASS  magnitudes ({\it Columns 5,
6, \&  7)}, LW2  and LW10 ISO  magnitudes ({\it  Column 8 \&  9}), the
apparent de--reddened bolometric  magnitude ({\it Column 10}), period,
amplitude, flag and quality parameter derived for the light--curve in
the MACHO--red  bandwidth ({\it Column 11,  12, 13 \& 14})  and in the
MACHO--blue bandwidth ({\it Column 15, 16, 17 \& 18}).

Values  of $99.99$  for the  magnitudes and  of $-99$  for  period and
     amplitude indicate missing data. Delmotte et al. \cite{nausicaa} 
have shown     that
     $J_{DCMC}=J_{2MASS}-(0.11\pm0.06)$                             and
     $K_{sDCMC}=K_{s2MASS}-(0.14\pm0.05)$ but we did not apply this 
correction to Table 4. 
     Amplitudes (A)  are the difference between
     the minimum and the maximum  value of the MACHO intensity and are
     expressed  in  $mag\times 100$.  In  a  handful  of sources  this
     calculation  might  be biased  by  spurious  data  points with  a
     magnitude excursion  compared to the sinusoidal  behaviour of the
     light--curve.  Perhaps  these points  are  real  and represent  a
     flare.  The period is in days.

\begin{table*}
\caption{Extract of the table of the properties of LPVs with ISO counterpart in the SMC}
\label{general}
\[
\begin{array}{lrrrrrrrrrrrrrrrrr}
\hline
\noalign{\smallskip}
\mathrm{ISO-MCMS}&\mathrm{I}&\mathrm{J}&\mathrm{Ks}&\mathrm{J}&\mathrm{H}&\mathrm{Ks}&\mathrm{LW2}&\mathrm{LW10}&\mathrm{m_{bol_0}}&\mathrm{P}&\mathrm{A}&\mathrm{f}&\mathrm{q}&\mathrm{P}&\mathrm{A}&\mathrm{f}&\mathrm{q}\\
 & \multicolumn{3}{c}{\mathrm{DCMC}} & \multicolumn{3}{c}{\mathrm{2MASS}} & \multicolumn{2}{c}{\mathrm{ISO}} & & \multicolumn{4}{c}{\mathrm{M.-RED}} & \multicolumn{4}{c}{\mathrm{M.-BLUE}} \\
J004844.4-732119 & 14.69 & 12.91 & 99.99 & 13.09 & 12.11 & 11.80 & 10.96 &  9.84 & 99.99 &  203 & 428 & 1 &  70 &  374 &  74 & 9 &  67 \\ 
J004844.5-731754 & 14.65 & 12.82 & 99.99 & 13.18 & 11.92 & 11.14 & 10.18 &  9.84 & 99.99 &  343 &  91 & 9 & 253 &  342 & 180 & 9 & 277 \\
J004844.7-732018 &  9.65 &  8.57 & 99.99 &  8.74 &  8.07 &  7.99 &  7.76 &  8.06 & 99.99 & 2900 & 401 & 5 &  20 & 2900 & 334 & 5 &  15 \\ 
J004849.4-731627 & 14.29 & 13.99 & 99.99 & 14.19 & 14.09 & 13.77 & 11.64 & 10.91 & 99.99 & 1900 &  95 & 5 &  46 & 2063 &  82 & 6 & 116 \\
J004849.8-732002 & 11.80 & 10.31 & 99.99 & 10.48 &  9.65 &  9.42 &  9.26 &  9.24 & 99.99 &   29 & 220 & 9 &  10 &  821 & 226 & 5 &  10 \\
J004850.7-732138 & 14.68 & 13.67 & 12.68 & 13.80 & 13.15 & 13.05 & 11.77 & 99.99 & 99.99 &  368 &  24 & 9 &  31 & 1667 &  42 & 5 &  11 \\ 
J004851.8-732239 & 10.90 &  9.45 &  8.41 &  9.63 &  8.83 &  8.57 &  8.04 &  7.52 & 11.21 &  352 & 456 & 9 &  46 &  350 & 301 & 9 &   7 \\ 
J004854.0-731557 & 13.62 & 12.37 & 99.99 & 12.59 & 11.85 & 11.69 & 11.58 & 99.99 & 99.99 &   10 &  19 & 5 &  68 &   10 &  21 & 5 &  19 \\ 
J004859.7-731155 & 13.04 & 11.80 & 99.99 & 11.96 & 11.17 & 11.00 & 10.74 & 99.99 & 99.99 &  368 & 184 & 9 &  21 &   21 &  29 & 5 &   4 \\ 
J004900.4-732224 & 14.16 & 12.47 & 99.99 & 12.65 & 11.70 & 11.35 & 10.74 & 99.99 & 99.99 &  239 &  67 & 1 & 137 &  237 &  99 & 2 & 207 \\ 
\noalign{\smallskip}
\hline
\end{array}
\]
\end{table*}

\subsection{Bolometric Magnitude} 
With the availability  of magnitudes from the  $I$ band to the
mid--IR band  LW10, it  is possible to  determine quite  precisely the
bolometric magnitude of the sources in the sample by integrating under
the spectral energy distribution (SED). This integration has been done
by fitting a spline to the  SED profile in the plane $(\nu, F_{\nu})$,
ex\-tra\-po\-la\-ting  linearly  to zero  flux  at  zero  frequency, and  also
extrapolating linearly at the other extreme of the SED through the two
near most points. The SED of all the sources detected by DENIS and ISO
were inspected one--by--one to exclude those which are either too red
or too  blue to describe  satisfactorily their SED with  the available
measurements. 

The adopted extrapolations  represent the SED very well,
almost  independently from  the  circumstellar o\-pa\-ci\-ty  for  all M  spectral
subtypes  (Loup et al.~{\it  in preparation}).   That article  gives a
detailed  determination of the  multi--band bolometric  cor\-rec\-tion for
AGB stars by combining data  from the major survey instruments: DENIS,
2MASS,  ESO,  IRAS, ISOCAM  and  MSX.   The  bolometric correction  is
obtained  from   the  different  combination  of  pass--bands  and  a
theoretical model  for the SED.  This is compared with  the correction
obtained from the integration under the  SED using a spline or a linear
fit through the flux points.  A fairly well sampled SED, that detects
most of the  energy emitted by the objects in  a set of cha\-rac\-te\-ri\-stic
bands, is  a necessary condition  to derive good
bolometric luminosity  ($M_{BOL}$) for  stars of different  M spectral
subclass and  C stars.  

The best  wavelength coverage for  the data in
the  sample  studied here  is  obtained  by  combining DENIS  and  ISO
magnitudes.  Though  2MASS $K_S$  magnitudes exist for  $455$ sources,
DCMC--$JK_S$  magnitudes exist  for $366$  sources and  DCMC--$IJ$ mag
exist for  about $450$ sources. On  one hand it would  seem natural to
combine $IJ$ DENIS magnitudes with 2MASS--$K_S$ magnitudes and the ISO
magnitudes, for the same  sources, to derive the bolometric magnitude.
On  the  other hand  the  DCMC--$IJK_S$  magnitudes  were obtained  by
simultaneous observations and  this is a unique constraint  to the SED
profile. 

Note that the amplitude of variation in the mid--IR bands can be as high 
as $1$ mag (e.g. van Loon et al. \cite{jacco98}, van Loon et al. 
\cite{jacco01}), this effect may influence the shape of the SED during the 
variability cycle. The  constraint of the $I$ band to  the shape of the
SED at  higher frequencies  is very important  for optically  thin AGB
stars  because the peak  of the  SED is  at about  $K_S$.  In  Loup et
al. ({\it  in preparation}) we  show that the linear  extrapolation to
zero flux at the longer frequencies  starting from the flux in the $J$
band subtends a  larger area than the extrapolation  starting from the
flux  in  the  $I$  band.  This effect  overestimates  the  bolometric
magnitude.   

Prior to  the  calculation  of  the  flux  emission  in  the
different bands and the integration  under the SED we de--reddened the
data adopting the extiction law by Glass (\cite{gext}).  For the DENIS
pass--bands $[A_V:A_I:A_J:A_{K_S}=1:0.592:0.256:0.089]$, and we assumed
zero  absorption  in  the  ISO  wave  bands.  We  used  $R_V=3.1$  and
$E(B-V)=0.065$, the latter is the  average of the measurements, in the
SMC, discussed in Westerlund (\cite{westbook}). The zero--point of the 
bolometric scale is computed with: $M_{BOL,\sun}=4.74$ and 
$f_{tot,\sun}=1.371\times 10^{-6}$ erg cm$^{-2}$ s$^{-1}$. A comparison 
of our bolometric magnitudes with the relations obtained by Alvarez et 
al. (\cite{alva}) and Montegriffo et al. (\cite{monte}) is given in 
Appendix B.

\begin{figure}
\resizebox{\hsize}{!}{\includegraphics{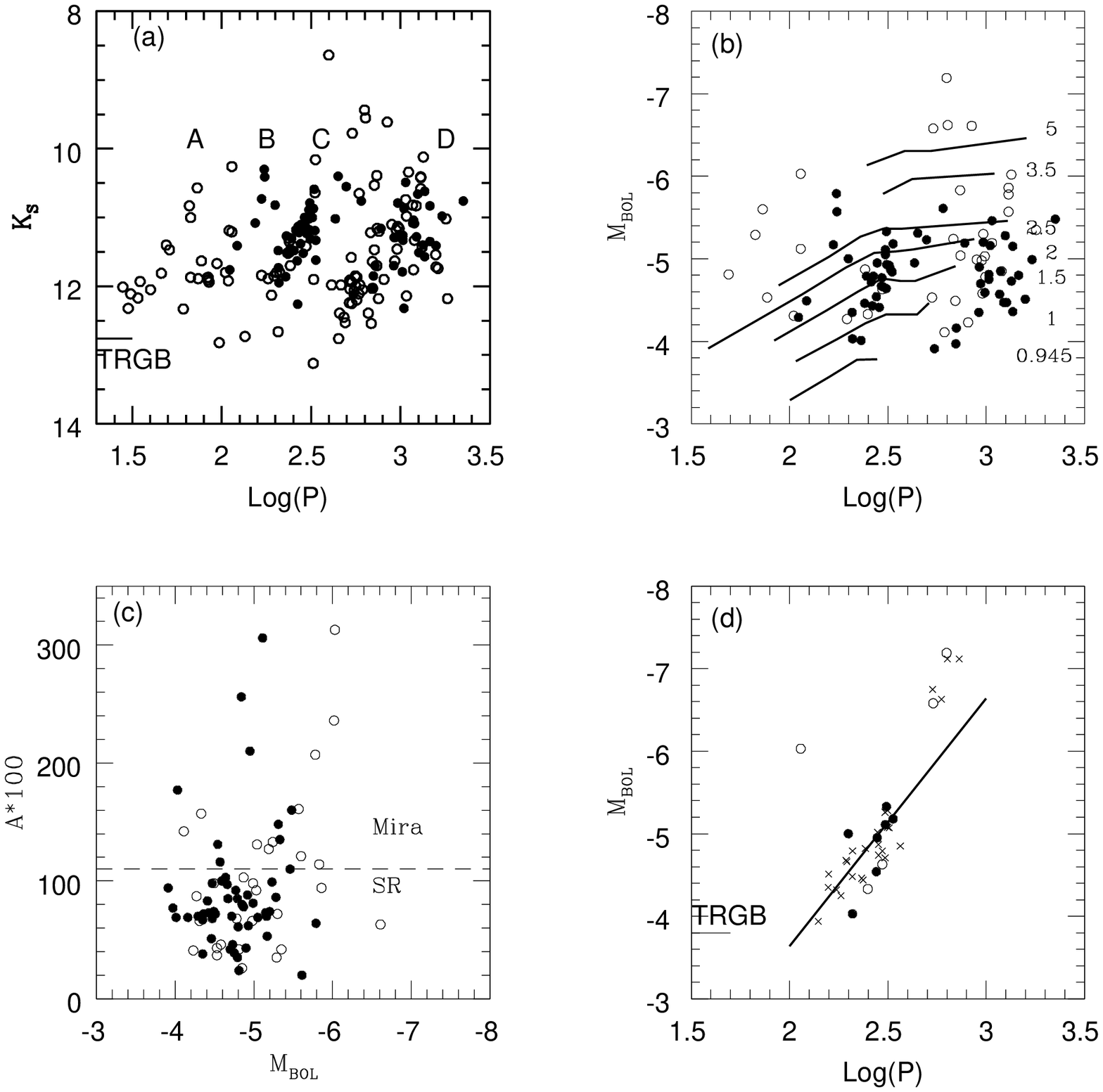}}
\caption{Diagrams for the sources in the sample with a MACHO--RED light--curve and 
$f=1$ or $f=6$. (a) $K_S$ versus $Log(P)$ relation. Letters ABCD identify 
the parallel sequences as defined in Wood et al. (\cite{woodiau}). Filled 
symbols indicate probable C--rich objects. (b) PLRs and overplotted 
theoretical tracks by Vassiliadis \& Wood (\cite{vass}). The numbers 
indicate the initial mass of the stars in each track. (c) Amplitude 
distribution as a function of bolometric magnitude. (d) PLRs for sources 
with an amplitude above the dashed line indicated in (c), $q>100$ and $f=1$.  
The continuous line is the Mira PLR fitted through the crosses corresponding 
to $P<420$ days obtained by Feast et al. (\cite{feastpl}); a distance 
modulus of $m-M=18.99$ (Cioni et al. \cite{cionitip}) has been applied. Symbols identify O--rich stars (open circles), C--rich stars (filled circles) and Feast et al. \cite{feastpl} variables (crosses).}
\label{plsmc}
\end{figure}

\section{Discussion}

\subsection{Period--luminosity relation(s)}
Fig. \ref{plsmc}a  shows the  relation  between the  logarithm of  the
period  (days) and  $K_S$.   Variable stars  are  distributed on  four
different parallel relations. These relations were first discovered by
Wood et  al. (\cite{woodiau}) from the  analysis of a  sample of MACHO
stars  in  the  LMC.   They   were  subsequently  found  by  Cioni  et
al.  (\cite{cionivar})  using  EROS2  (Lasserre  et  al.  \cite{lass})
light--curves, by Noda et al.  (\cite{moa}) using the MOA database and
by  Lebzelter et al.  (\cite{agape}) using  the AGAPEROS  data.  These
latter  measurements refer  also to  samples of  stars in  the  LMC. A
preliminary comparison between the relations in the LMC and in the SMC
has  been shown  by Cioni  et al.  (\cite{cionijap}),  concluding that
sources  in both  galaxies occupy  the same  relations in  the diagram
($K_S$, $Log(P)$).  

There are a few differences that depend on the way
the two samples  were selected and the light--curves  extracted: in the
SMC there are sources with longer periods because the MACHO data cover
a  longer time  range than  the  EROS2 data.  Thus it  is possible  to
identify longer  periodicities.  Also there  is a relative lack of  SMC sources
around the tip of the red  giant branch (TRGB) because the LMC sources
were selected  from the DCMC and  not among those  sources detected at
$7$ and $12$ $\mu$m. It may be  that AGB stars that are not yet in the
TP--AGB phase or are in a  very early phase of their TP--AGB evolution
have  a mid--IR  emission  below the  ISO  detection limit. The 
ISO survey was indeed designed to find mass--loss, not to detect sources 
without circumstellar extiction.  

The
analysis of the light--curves of  LMC sources detected by ISO has just
started.  
The $K_S$ magnitudes are always a one epoch measurement, and
coupled with the amplitude of variation, that in the $K_S$ band can be
as  high as  $1.5$  mag for OH/IR  stars, prevent  us  from de\-tec\-ting  a
difference (if any)  in the width of the PLRs  due to metallicity. The
same   situation   holds   for   stars   in  the   Bulge   (Alard   et
al. \cite{alard}),  which have  the additional  disadvantage  that the
distance to  the AGB  stars is  too uncertain to  find all  four PLRs.
Furthermore, AGB stars in stellar  clusters are too few to outline the
presence of one  and possibly two PLRs. There is  evidence that in the
 elliptical galaxy NGC5128 one  and maybe two  PLRs exits for
LPVs (Rejkuba \cite{marina}).

The atmosphere of AGB stars is dominated by molecules of TiO and VO if
the star is O--rich (C/O$<1$) or by molecules of CN, C$_2$ and other 
carbonaceous molecules if the star
is  C--rich  (C/O$>1$).   Stars  of  both types  are  easily
statistically  distinguished in  the $(J-K_S,  K_S$) colour--magnitude
diagram (Fig. \ref{spectra}a) because of the different strength of the
molecular absorption  bands at these particular  wavelengths (Cioni et
al. \cite{cionivar}, Loup  et al. \cite{loupbmb}).  The  dividing line
at $(J-K_s)=1.33$ takes into account the shifts calculated by Delmotte
et  al.~(\cite{nausicaa}) between  2MASS and  DCMC magnitudes  and has
been defined  in Cioni  \& Habing (\cite{cionicm})  for the  whole AGB
population  of the  SMC.  It  was also  formerly checked  by  Cioni et
al. (\cite{cionitip})  via the  cross--identification  with the  large
sample of  spectroscopically identified  C--rich stars of  Rebeirot et
al. (\cite{reb}).  Filled symbols in Fig. \ref{plsmc} indicate C--rich
objects. In Fig. \ref{plsmc}b they define relation B and C, and dominate 
relation D (possibly offset from the oxygen--rich stars).

\subsection{Mass and Age}
According to  the Vassiliadis and Wood  (\cite{vass}) prescription for
the evolution  of low-- and intermediate--mass stars  carbon stars are
less  massive than  about  $3  M_{\sun}$.  
  They  form  after  a certain  number  of  shell
flashes when the atoms of carbon overnumber the oxygen atoms. 
Hot  bottom burning  may  prevent massive
stars from becoming C--rich. Each  track drawn in Fig. \ref{plsmc}b starts at
the  first shell--flash  cycle  on  the TP--AGB  phase.   Most of  the
C--rich  stars in  the  SMC sample  have  a mass  between  $1$ and  $3
M_{\sun}$. The brightest stars ($M_{BOL}<-7$) are probably supergiants
(Wood et al. \cite{woodsg}). The most massive AGB stars with a mass
of about  $5 M_{\sun}$  are about $0.1$  Gyr old. The  oldest detected
AGB stars are  about $9$ Gyr old, but  most of the sample is from
$0.6$ to $2$ Gyr old.

In the  LMC (Cioni  et al. \cite{cionivar})  C--rich stars  occupy only the
brightest part of  the PLRs, thus in a  lower metallicity environment,
such as  that of the SMC,  C--rich stars form at  lower masses, unless
these stars  are in the  luminosity dip that follows  a thermal pulse or  
at the
minimum  of  their   light--curve.   Menzies  et  al. (\cite{menzies})
recently showed  that the top magnitude  of the AGB stars  in $K_S$ in
the  dwarf   galaxy  Leo  I ([Fe/H]=-2.0)  are  mostly  carbon  stars.   
They  have
$J-K_S>1.1$.  The  value of  the  $J-K_S$  that di\-scri\-mi\-na\-tes  between
O--rich and C--rich stars is  a function of metallicity. The lower the
metallicity,  the  bluer the  colour.   Obscured  AGB star  candidates
($J-K_S>2$)  might  also be  of  C type ($50\%$ level) 
and  the  observation at  two
different epochs indicates that at least one of these object might be a
Mira variable ($\Delta K_S=0.87$).

\subsection{Large Amplitude Variables}
Fig.~\ref{plsmc}c shows the amplitude of the light--curves as
a  function of  $M_{BOL}$ for  C--rich and  O--rich  objects.  
The sources with $A*100>110$ and $f=1$ have $q>100$ and are 
distributed in    the   PLR(s)   as    shown    in
Fig.~\ref{plsmc}d.  
  They occupy relation  C which  is the  location of
Mira  variables  in the  LMC  and the  Galaxy.   Over  plotted is  the
sequence derived by Feast et al. (\cite{feastpl}) and the single 
LPVs (crosses) as re--analysed in this work (Appendix A). This sets the 
minimum level of the quality parameter for Mira variables. 
This  sequence  was  obtained  including  both
O--rich and C--rich  AGB stars with a period  $P<420$ days. 
A distance modulus of $18.99$ has been used (Cioni et al. \cite{cionitip}). 
The single
carbon  at  $Log(P)=2.1$ and  $M_{BOL}=-6$ is of large amplitude 
 because of flare--like events super imposed on a regular variation 
of smaller amplitude, thus this source should not be regarded as a Mira 
star. 

 The scatter in the PLR, of the sources in this study,  
results from by the  unavailability  of bolometric light--curves. 
However, it is consistent with the sigma associated with the relation 
derived by Feast et al. \cite{feastpl}. 
Estimating the magnitude of this effect would be valuable for 
studies that plan to use AGB stars as tracers of the 
star formation history of systems as far as $250$ kpc 
(Ku\^{c}inskas et al. \cite{kucin}). This preliminary investigation on 
the use of AGB stars concerns mainly early--type AGB stars which vary with 
amplitudes much below that of Mira--type stars. On the other hand 
Ku\^{c}inskas et al. combined broad--band observations from the literature. 
It should not be neglected that the amplitude of variability of AGB stars  
is different in each photometric band. Only simultaneous observations 
in different bands, as those 
that will be available with GAIA, provide a coherent SED from which a 
single--epoch bolometric magnitude can be derived. 

Finally we conclude that a large amplitude and a large value
of $q$  indicate a clear and  regular light--curve; thus  a Mira star.
There are some sources of large amplitude and with 
a large value of the quality parameter suspected of multi--periodicity. 
Their main period does not correspond to relation C. 
 A few  Mira variables in  sequence D
have been found by Alard  et al. (\cite{alard}). On the contrary, this
PLR could be populated by binary  stars with an AGB companion (Wood et
al.~\cite{woodiau}).  In what follows we mainly discuss the properties of the
stars with a light--curve in the red band of MACHO 
and $f=1$ or $f=6$.

\begin{figure}
\resizebox{\hsize}{!}{\includegraphics{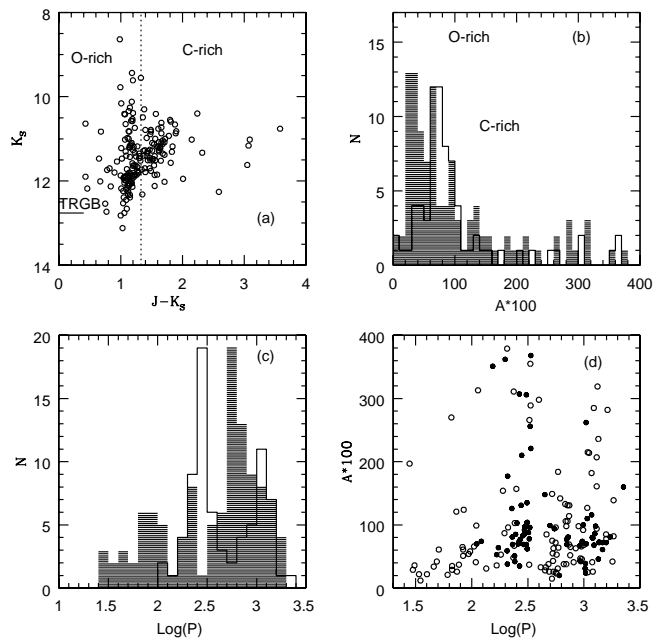}}
\caption{Variable O-- and C--rich AGB stars (a). The histogram of the amplitude of variation (b) and that of the period distribution (c) are shown for sources of both types. The amplitude as a function of period is shown in (d).}
\label{spectra}
\end{figure}

\subsection{Mira \& Semi--Regular Variables}
The distinction  between Mira and Semi--Regular (SR)  stars 
has become more  and  more  subtle with  the
a\-vai\-la\-bi\-li\-ty  of  more  sensitive  long term  observations  and  
large
samples of stars.  The stellar  light--curve can be as regular in SR 
as   in  Mira  stars.    Lebzelter  et   al.~(\cite{agape})  define   a
classification  criterium for  LPVs based  only on  the  re\-gu\-la\-ri\-ty 
and
multi--periodicity of  the light--curves. Most of the  Mira stars fall
into the class of regular variables but not all regular variables have
large amplitudes of variation. SR belong to a second class and have on
average a smaller  amplitude than the stars in  the first class.  Into
the third  class fall  irregular variables, that  do not show  a clear
periodicity.  The  Mira phase is believed  to be just  a moment during
the AGB  evolution and  the star  can thus become  or have  been SR. 
This scenario  is supported by observations of very nearby variable  stars 
over  a  period of  about  $90$  years  (i.e. Percy  \&  Bagby
\cite{perbag}, Kiss et al. \cite{kiss}).
However the evidence, e.g. from globular clusters, suggests that most 
SRs are in the pre--Mira stage.

Mira variables  lie on  one PLR (Feast  et al. \cite{feastpl})  and SR
occupy  this and  other PLRs  according to  their mode  of pulsation
(Wood  et  al. \cite{woodiau}, Cioni  et  al. \cite{cionivar}, Noda  et
al. \cite{moa}, Lebzelter et  al. \cite{agape}). If Mira variables are
selected as  discussed in Sect. 4.3 then they are
clearly concentrated on the C  PLR. This selection criterium is based
on the  amplitude of variation. However, the sources are the 
most  regular as well ($q>100$).

Whitelock  et al. (\cite{wfmo}),  by  analysing the  light--curves of  a
dozen  C-rich AGB  stars in  the Milky Way Galaxy,  
showed that  the distinction
between  Mira and  SR va\-ria\-bles  is not  necessarily very  clear among
carbon stars,  and may not  be useful  as a dis\-tinc\-tion  for C--rich 
stars  as  it is  for  O--rich  stars.  Fig. \ref{spectra}b shows  the
histogram of the distribution of the amplitude of variation separately
for O--rich and C--rich stars.  Perhaps there is a distinction between
Mira, the pulsators at the  tail of the amplitude distribution(s), and
SR. Both O--rich and C--rich sources 
 have a histogram with a dominant peak at about
$A\times100=40$   and  $A\times100=80$, 
respectively, and a tail to  larger amplitudes.  The fact that C--rich
stars  have  on  average  larger  amplitudes  than  O--rich  stars  is
different from what  was found in the LMC.  There, both histograms had
the same distribution (Fig. 4 in Cioni et al. (\cite{cionivar})).
The same histogram but for the amplitudes derived from the BLUE light--curve 
shows that O--rich stars peak at about $A*100=50$ and C--rich stars at 
about $A*100=100$. This confirms that the average amplitude is larger 
in the blue than in the red spectral domain.

\subsection{Periodicity peaks}
The histogram of the distribution of the variable sources versus their
period is  shown in Fig. \ref{spectra}c.  The  distribution of O--rich
stars has a main peak at  roughly $Log(P)=2.80$ and two lower peaks at
about $Log(P)=1.95$ and $Log(P)=2.30$. C--rich stars have well defined
peaks at  about $Log(P)=2.45$  and $Log(P)=3$. Note  that most  of the
sources  with  $2.53<Log(P)<2.59$  have  been excluded  because  their
period determination is suspected of aliasing.

Fig. 4 in Cioni et  al. (\cite{cionivar}) shows that O--rich AGB stars
in the LMC have a strong broad peak between $Log(P)=1.5$ and $2.0$, and a lower
peak at about $Log(P)=2.45$, while C--rich AGB stars are homogeneously
distributed from $Log(P)=1.5$ to $2.5$.  The first LMC peak of O--rich
AGBs  is probably  a consequence  of  the selection  criterium used  to
define the two samples. The  SMC sample is extracted from sources with
an emission  in the mid--IR, these  stars are  
brighter and  with longer  periods than the  stars in the  LMC sample
which were selected from the DCMC database. The difference in strength
is  probably due  to  the mean  difference  in the  AGB  type
between the two  Clouds. There are more carbon stars  in the SMC which
is  on  average  more  metal  poor  than  the  LMC  (Cioni  \&  Habing
\cite{cionicm}).   This  effect  is  also important  for  the  sources
populating relation C which are predominantly C--rich in the SMC and
  associated to  the peak  at about  $Log(P)=2.45$. 
That the majority of accepted Miras in the SMC are carbon stars was first 
re\-co\-gni\-zed by Lloyd Evans et al. (\cite{llglca}) as the result of a 
search 
for large--amplitude red variables in the Radcliffe Variable Star Field. 
The  longer time
range spanned by the MACHO observations allows to detect periodicities
above $Log(P)=2.8$ which was about  the limit reached in the LMC study
using EROS light--curves. The large  number of sources detected in the
SMC  at  these  periods  are  likely  to  be  multi--periodic.   Their
secondary period will be part  of relation A, B or C, thus they
will  enhance the  number of  O--rich stars  between  $Log(P)=1.5$ and
$2.5$. In more detail we found that in our sample, among those with a red 
light--curve and $q>15$, $24$ sources have $flag=6$. 
These sources have $2.0<Log(P)<2.3$ which, taking into account the 
differences between the various period determinations and the mean,
 gives a period ratio from $1.5$ to $2$. 

The   same    sample   of   LMC    stars   analysed   by    Cioni   et
al.   (\cite{cionivar})  was  afterwards   studied  by   Lebzelter  et
al. (\cite{agape})  using the  AGAPEROS datasets.  Ninety per cent  of their
sources  were classified into:  regular, SR,  irregular and
other variability classes. Their analysis  covers a time window of up to
$900$ days and  is thus incomplete for the longest  periods, but it is
more complete for  sources with an amplitude of  variation below $0.1$
mag.  The  authors  found  that  the period  distribution  of  regular
variables  has  two  peaks:  one  at $Log(P)=2.0$  and  the  other  at
$Log(P)=2.5$.  The  latter  is  probably  formed by  Mira  stars.   SR
variables  peak  at about  $Log(P)=1.8$.  The  peak  at the  shortest
periods is probably the same as the one found here. The peak of mostly
Mira stars corresponds to the location  of the peak of the C--rich SMC
stars.  In  the Baade's window, Glass \&  Schultheis (\cite{gs}) found
that most of the late--type M giants have $Log(P)<2.0$; those few with
a secondary longer period  have about $Log(P)>2.4$. The histogram that
can be  derived from their Fig.~2  is very similar to  that of O--rich
stars in the SMC (Fig. \ref{spectra}c).

\subsection{($J-K_S$, $K_S$) diagram} 
\begin{figure}
\resizebox{\hsize}{!}{\includegraphics{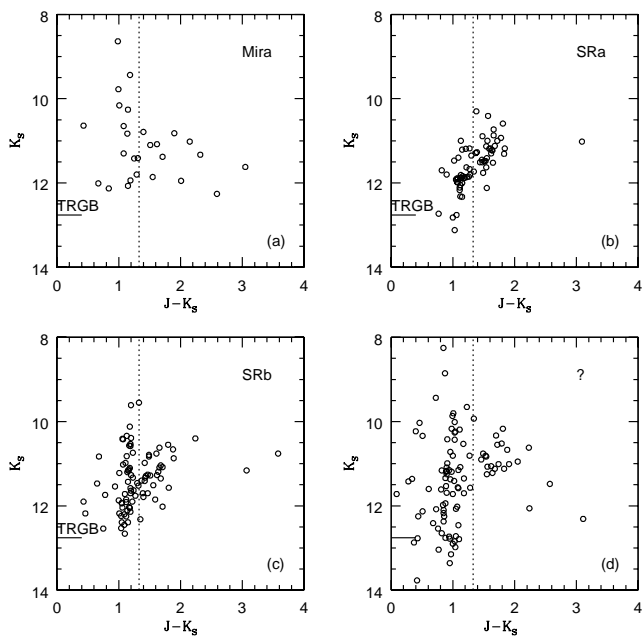}}
\caption{Colour--magnitude diagrams ($J-K_S$, $K_S$) for variable stars 
classified as Mira (a), SRa (b) and SRb (c) according to their 
periodicity and  amplitude of variation. (d) is the same 
diagram for variable sources which could not be classified in one of the 
three categories because of low S/N or an extremely irregular light--curve. 
The vertical dotted line indicates the separation between O--rich and C--rich 
stars as in Fig. \ref{plsmc}.}
\label{varia}
\end{figure}
In  a  similar  fashion  to   the  LMC  sample  (Fig.~7  in  Cioni  et
al. \cite{cionivar}),  variable  stars  of  a different type and
va\-ria\-bi\-li\-ty class are plotted  in the colour--magnitude diagrams 
(CMDs)
in Fig. \ref{varia}.   The distinction has been made  as follows: Miras 
have  $A\times100>110$  independently of their type,  
and a regular light--curve with a well defined single periodicity. 
Similar sources but with a smaller amplitude of variation are SRa stars.  
SRbs  are those stars  which indicate the
presence of a secondary period.  Note that in this last category there
is no distinction based on the amplitude of variation.  As in the LMC,
Mira stars  are uniformly distributed  in the regions  where optically
thin O--rich  and C--rich,  and also more  optically thick  stars, are
located   (Fig. \ref{varia}a).  SRa   and  SRb   are   almost  equally
distributed within  the two branches of O--rich  and C--rich optically
thin  AGB  stars.  Fig.~\ref{varia}d  shows  the  distribution of  the
sources that  could not be classified  in one of  the three categories
either because the  light--curve has a too low S/N  or it is extremely
irregular.  Most of these stars ($180$ out of $194$) have $A>10$, thus
they are variable  stars and lie above the  TRGB, $K_S=12.62$ (Cioni et
al. \cite{cionitip}). It is  possible that stars considerably brighter
than  $K_S=10$  are  supergiants  and  that  fainter  stars  at  about
$(J-K_S)=0.4$ are  foreground objects. Schultheis  et al. (\cite{sch})
found va\-ria\-ble star candidates at  $(J-K_S)<1$. They conclude that it is
probably a result of uncertainties in the  determination of extinction.
However,  they find  a displacement  of  the Bulge  stars detected  as
variables in  both $J$  and $K_S$ to  bluer colours than  the sequence
indicated  by  all  detected  ISO  stars in  the  diagram  ($LW2-LW3$,
$LW3$). It  would be interesting  to know if  some of these  stars are
supergiants and of which type.

Finally we  conclude that all the  sources in the  sample are variable
stars because they have amplitudes  above or equal to $0.1$ mag either
in the  RED or  BLUE wave  band. A similar  conclusion was  reached by
Alard et al. (\cite{alard}) for a  sample of sources in Baade's window
detected  at  $7$  and  $15$  $\mu$m.  The  emission  at  the  longest
wavelength is  a signature  of 
mass--loss  rate. Using  their  dependence between  $M_{BOL}$ and  the
mass--loss rate and assuming that the mass--loss rate of highly 
evolved AGB stars does not depend very strongly on metallicity 
(van Loon \cite{jacco}), we derive that 
the sources in the present sample lose mass at a rate
from    $10^{-7}   M_{\sun}yr^{-1}$    ($M_{BOL}=-4$)    to   $10^{-4}
M_{\sun}yr^{-1}$  ($M_{BOL}=-7$),   but  most  of   the  objects  have
$M_{BOL}=-5$ which corresponds to  a mass--loss rate of about $10^{-6}
M_{\sun}yr^{-1}$. Note that a complete study of the dependence of mass--loss 
includes temperature, mass and outflow velocity as well.


\subsection{($K_S-LW10$, $LW10$) diagram}

The $7 \mu$m LW2 filter does not include a large dust contribution for
non--Mira  sources and  it is  mostly  a measure  of the  photospheric
emission.  On  the  other  hand,  for stars  with  a  relatively  thick
circumstellar  dust envelope,  this band  is strongly  affected  by the
presence  of dust  (Alard et  al.~\cite{alard}, Omont et al. \cite{omont}).  
The  $12  \mu$m LW10
filter, which is very similar to the IRAS $12 \mu$m band, includes the
$9.7 \mu$m  silicate and  the $11.3 \mu$m  SiC dust features.  Thus the
colour ($K-LW10$)  provides a strong indication of the
amount of circumstellar dust for the entire sample.

\begin{figure}
\resizebox{\hsize}{!}{\includegraphics{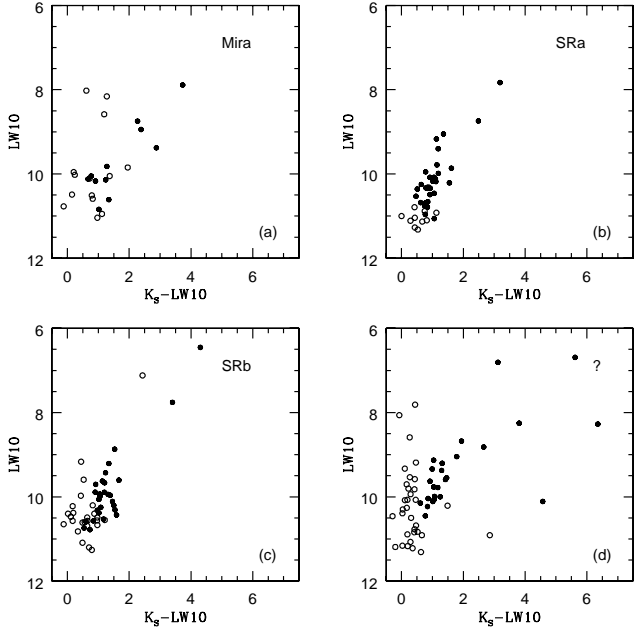}}
\caption{Colour--magnitude diagrams ($K_S-LW10$, $LW10$) for the same groups  
of sources as in Fig.~\ref{varia}.}
\label{cmd2}
\end{figure}

Fig.~\ref{cmd2}  shows  the   colour  magnitude  diagram  ($K_S-LW10$,
$LW10$) for  the sources in the present  sample, distinguishing between
Mira  and  SR  variables.  Except for  a  few  SR  all  the  sources  with
$(K_S-LW10)>2$  are  Mira   variables  (Fig. \ref{cmd2}a).   From  the
location  of  some  K--  and  M--type supergiant  (SG)  stars  in  the
ISO--MCMS  one derives  that  they have  $(K_S-LW10)<1$ and  $LW10=8$.
They  follow a  sequence  that extends  almost  vertically to  fainter
magnitudes,  but the  nature of  those sources  with $LW10>10$  is not
confirmed.    Bright  O--rich LPVs
would have $(K_S-LW10)=3$ and $LW10=7$.  These stars would have a late M
subtype and not  many of them are expected in  the SMC. A quantitative
statement will be possible once the  study of the LMC LPVs detected by
ISO  is  completed.   A  few  SRb  stars  populate  the  fainter  part
(Fig.~\ref{cmd2}b, c). There are many supergiant stars along the whole
sequence  that were  not classified  in  any of  the three  categories
(Fig. \ref{cmd2}d).  It is thus not surprising that their light--curve
is   more  difficult   to  classify   because   of  a   low  S/N   and
irregularities. Thus,  except for a  few stars overlapping  the region
occupied by SR and Mira stars we are able to distinguish sources with a poor
or  non--regular  light  curve.   A  more  careful  look  at  the near--IR
colour--magnitude   dia\-gram  (Fig. \ref{varia})   confirms   that  the
unclassified stars with $(J-K_S)=1$ or a bluer colour are experiencing
the supergiant phase because  AGB stars have overall $(J-K_S)>1$. This
is quite  a good criterion to discriminate  between supergiant and
AGB stars, though  the distinction is better achieved  in the combined
near-- and mid-- infrared diagram (Fig. \ref{cmd2}).
 
Most carbon  stars  are located  at  about  $(K_S-LW10)=1$ and  $8<LW10<11$
(Fig. \ref{cmd2}--filled  points);  the   sequence  of  dust  obscured
C--type stars extends to redder colours but the $LW10$ mag follows a
shallower slope such as $(K_S-LW10)=4$ at $LW10=8$. On average C--rich
stars are redder than O--rich stars.

Planetary  nebulae and  emission  line objects, included in the ISO--MCMS 
catalogue, also  have red  colours
($K_S-LW10>2$) but  they have fainter magnitudes with  respect to
Mira stars  and are roughly located  in a parallel  sequence. Three of
these objects are present in Fig. \ref{cmd2}d.
The object at $LW10<8.5$ and $K_S-LW10>4$ is a known carbon star while 
the object with the same $LW10$ magnitude and a bluer colour is 
a known variable M giant, because of its red colour we 
erroneously classified it as a carbon star.

\subsection{($K_S-LW2$, $LW2-LW10$) diagram}

\begin{figure}
\resizebox{\hsize}{!}{\includegraphics{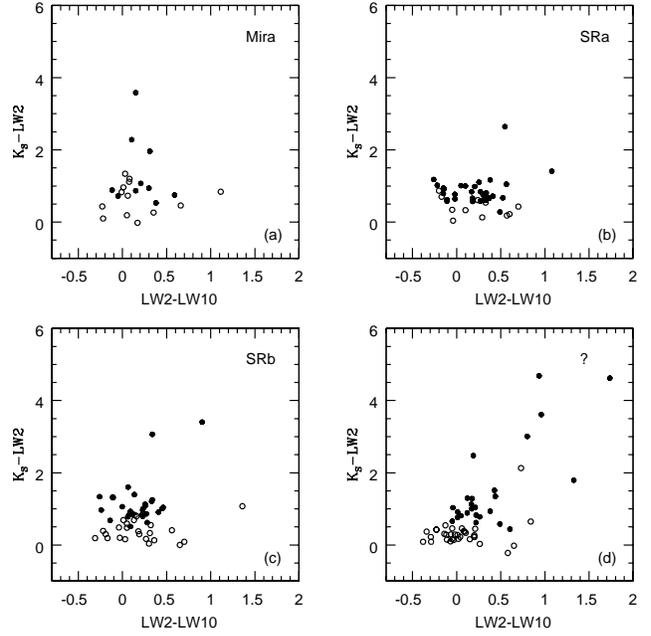}}
\caption{Colour--colour diagrams ($K_S-LW2$, $LW2-LW10$) for the same groups 
of sources as in Fig.~\ref{varia}.}
\label{cmd3}
\end{figure}

Fig.~\ref{cmd3}  shows  the colour--colour  diagram  of the  variables
 identified in  this study.   The O--rich sources  with $(LW2-LW10)>0.4$
 are AGB stars or red supergiants.  These AGB stars have a significant 
circumstellar extinction. This  is also the colour of obscured O--rich
 and  C--rich   stars,  and  those  with  the   highest  amplitude  of
 variability have $(K_S-LW2)>1.5$.  O--rich stars without a considerable
 mass loss   have    approximately   $(LW2-LW10)=0$   but    can   have
 $0<(K_S-LW2)<1$. There is not  a remarkable distinction between SRa and
 SRb pulsators.  At  about $(K_S-LW2)>3.5$ one can find  PN and emission
 line objects.

It seems (Fig. \ref{cmd3}d) 
that we  may have  missed the classification  of a  few Mira
variables, O--rich  and C--rich SR stars.   At this stage  it is worth
going back  to the  specific light--curves to  find out the  reason of
this  missed  classification and  to  perhaps  identify some  peculiar
objects.  Among  the  objects  with $K_S-LW2>1.5$  (Fig.  \ref{cmd3}d)
J005113.6$-$731035,  a  known  carbon  star  (RAW 658),  has  a flat
light-curve, while the light--curve of J005136.5$-$732016, identified as
a PN,  is irregular. Another PN  is J005157.9$-$731421 which  may have a
re\-gu\-lar  light--curve  with  an  amplitude  of about  $0.8$  mag.  The
periodogram of the other objects, J004849.4$-$731627 (emission line
star),  J005212.9$-$730852  (variable  M  giant) J005304.7$-$730409 and 
J010020.6$-$730648 (BMB-W  29 carbon star)  shows a  very
low amplitude with flare--like events.

\subsection{Miscellaneous}
SRb, as have  been defined in this work, do  not constitute a separate
class of variables. They  are clearly multi--periodic objects and from
their photospheric and dust colours they do occupy the same regions of
SRa and Mira  stars. Until a confirmation of  the specific theoretical
models (such as  those developed by Winters et  al. (\cite{winter}) to
explain  the  multi--periodicity  of  the  light--curve  variation  by
coupling of  stellar pulsation with  the dust formation) or  follow up
radial  velocity  observations (to  infer  the  binary  nature of  the
sources)  are available, the  most appropriate  class to  which assign
these stars is that of general SR variables. This is also strengthened
by  the fact  that most  of  the objects  in the  sample indicate  the
presence of  more than one periodicity.  A sophisticated observational
and  theoretical  analysis of  the  multiple  periods  is required  to
further address this point. Because we did not discriminate on the 
amplitude of SRb stars some of them do have Mira--like amplitude but are 
likely to have more than one significant period.

\begin{figure}
\resizebox{\hsize}{!}{\includegraphics{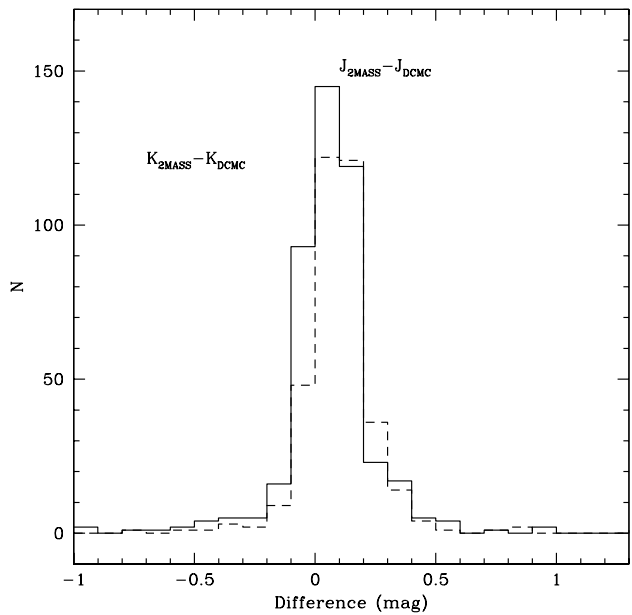}}
\caption{Histogram of the differences between the 2MASS and DCMC magnitudes 
of all the sources in the sample.}
\label{irvar}
\end{figure}

Most  of the  DENIS  sources in  the  present sample  were observed  in
December  1996  while  2MASS   observations  date  August  1998.  This
difference  in time  of  about two  years  may allow  us to  recognize
variable stars by  com\-pa\-ring their detection in the  near--IR bands of
DENIS  and 2MASS.  Fig. \ref{irvar}  shows the  distribution  of these
differences for  all the  sources in the  sample. We checked  that the
separate histogram  of sources with  an assigned MACHO  periodicity is
the same as  that of sources with Flag$=5$ (a  period was not assigned
but the source is probably a variable). The histogram does not account for the
systematic  difference  between  the  2MASS  and  DCMC  magnitudes  as
measured  by  Delmotte  et  al. (\cite{nausicaa}).  Selecting  only  the
variable  stars  with  the   largest  amplitude  produces  a  si\-mi\-lar
histogram. The displacement  of the peak from zero  indicates that the
sources are variable stars. Note that the difference  is approximately the
same in the $J$ and $K_S$ bands.


\section{Conclusions}
The  stars selected from  the ISO--MCMS  with a  confident DENIS/2MASS
counterpart have  been cross--identified with the  MACHO database. The
extracted light--curves  have been analysed to derive  the main period
and  amplitude  of  pulsation.   The  bolometric  magnitude  has  been
calculated by  integrating the  spectral energy distribution  from the
$I$ DENIS to  the $LW10$ ISO wave band. This,  in combination with the
correlation of  Alard et al. (\cite{alard}) between  $M_{BOL}$ and the
mass--loss  rate, indicates  that most  of  the stars  in the  present
sample lose $10^{-6}M_{\sun}yr^{-1}$ of matter.  A distinction between
Mira and  SR is obtained  by applying an amplitude and regularity selection
criterion. The histogram of the amplitude
of  pulsation of O--rich and C--rich  stars has  a similar
distribution,   but   on  average   C--rich   stars   have  a   larger
amplitude. This may indicate that  either most of the C--rich stars in
the  SMC  are  of Mira  type,  or  that  the metallicity  affects  the
amplitude in  such a way that  in a lower  metallicity environment the
amplitude of  pulsation is larger.  This effect cannot yet  be checked
for in other  metal--poor galaxies in the  Local Group, because despite
the fact that many AGB stars  have been discovered there is not enough
information on their variability and type.

The  most obscured  stars have large  ($K_S-LW2$)  and ($LW2-LW10$)
colours.  This indicates sources with a thicker circumstellar envelope
and allows to  easily distinguish AGB stars from  supergiant stars and
other classes  of objects  such as PN  and emission line  sources. The
dif\-fe\-ren\-ce between the 2MASS and DCMC magnitudes suggests that most of
the sources in the sample are variables.

\begin{acknowledgements}
The first  author is very  grateful to J.-P.  Beaulieu for  providing the
algorithm to  detect the  pe\-rio\-di\-ci\-ties.  This paper  utilises public
domain data  obtained by the MACHO  Project, jointly funded  by the US
Department of  Energy through  the University of  California, Lawrence
Livermore  National Laboratory under  contract No.   W-7405-Eng-48, by
the  National  Science  Foundation  through the  Center  for  Particle
Astrophysics  of  the   University  of  California  under  cooperative
agreement  AST-8809616, and  by the  Mount Stromlo  and  Siding Spring
Observatory, part of the Australian National University. This publication 
makes use of data products from the Two Micron All Sky Survey, which is a 
joint project of the University of Massachusetts and the Infrared Processing 
and Analysis center/California Institute of Technology, funded by the 
National Aeronautics and Space Administration and the Mational Science 
Foundation.
\end{acknowledgements}

\appendix

\section{The period--luminosity relation of Mira stars}
In order to improve our understanding of the identification of Mira--type 
variables we decided to re-analyse the LPVs studied by Feast et al. 
\cite{feastpl}. These are the Mira stars that define the 
period--luminosity relation in the LMC. 
We could find the B1950 coordinates of $42$ out of the $55$ sources 
with a ``good'' or ``fair'' light--curve as classified by Feast et al. 
(\cite{feastpl}).  
We could not find in the literature the coordinates of the objects 
from the sample of 
Glass \& Lloyd Evans (\cite{glallo}) and 
for three objects from the list of Glass \& Reid (\cite{gr1985}) or 
Reid et al. (\cite{reidetal1988}). However, there still remain 
enough sources for our purpose. 

We searched in the MACHO database for the light--curve of the closest LPV 
to these $42$ sources. The identification was easy in most of the cases 
because of the highly regular and large amplitude light--curve typical 
of a Mira star.  Five sources (W158, GR28, GR4, RGC91, RGC89)\footnote{The 
designation of the sources follows that of Feast et al. \cite{feastpl}: 
W=Wood et al. \cite{wood1985}, GR=Glass \& Reid \cite{gr1985} and we 
introduce RGC=Reid et al. \cite{reidetal1988}.} are  
unfortunately outside the fields observed by MACHO. For one source the MACHO 
web site gives a persistent initialization error (W94) 
and for four additional sources (GR13, GR21, RGC37, RGC53) 
 we could not find a nearby counterpart (a first possible candidate is further 
than $8$\arcsec). Only one source (W30) has two possible 
MACHO counterparts and due to the low accuracy of the initial coordinates 
compared to the MACHO coordinates we decided to remove this source from 
the discussion. 
Finally we could successfully extract the light--curves of $31$ sources.

We determined the main periodicity, as described in Sect. 3, of the sources 
with a MACHO--red light--curve. 
We obtained periods that are at most $20$ days different from those derived 
by Feast et al. (\cite{feastpl}) except for $4$ sources which result to 
have a period coincident with one of the extremes of the searching window  
(GR2, GR3, GR18 and GR29, they have been removed from the following analysis) and 
source GR17 for which we derive a period of $51$ days shorter. The period of 
source RGC20 falls in the region of aliases (Sec. 3), thus the source has been 
removed from the following discussion.
The quality parameter of the $26$ sources with a convincing period determination 
is always above $100$ and the difference between the assigned 
period and the mean of all the determinations with $q>15$ is of the order 
of $1$ day for most of the sources but always below $10$ days. 
These numbers may be different in the case of MACHO--blue light--curves. 

Table \ref{feasttab} lists the name of the sources ({\it Column 1}), the 
coordinates at the epoch J2000 precessed from the original coordinates 
({\it Columns 2 \& 3}), the MACHO 
identification code ({\it Columns 4, 5 \& 6}), the distance in arcsec 
between the Feast et al. (\cite{feastpl}) 
source and the MACHO counterpart ({\it Column 7}), 
the period, amplitude and the 
value of the quality parameter for the red light--curve ({\it Columns 8, 9 \& 10}). 

We conclude that by selecting, from our sample, large amplitude 
sources with a single periodicity and $q>100$ we are identifying Mira variables 
of the same type as those studied by Feast et al. \cite{feastpl}. They define 
a tight PLR and the small scatter in Fig. \ref{plsmc}d is probably due 
to the a\-vai\-la\-bi\-li\-ty of only one epoch bolometric magnitude. Note that with 
this selection criterion we also find sources at very bright magnitudes 
($M_{BOL}=-7$) which deviate from the extrapolation of the PLR defined only 
for sources with $P<420$ days. We estimate from Fig. \ref{plsmc}b that these 
stars have $M>5M_{\sun}$ and as previously noted by Feast et al. 
\cite{feastpl} the PLR breaks at high masses. These stars are likely to be 
``over--luminous'' as a result of Hot Bottom Burning (Zijlstra et al. 
\cite{albi}, Whitelock et al. \cite{whiti}). However, because of their very 
high luminosity they could also be supergiants.

\begin{table*}
\caption{Table of MACHO counterparts of Feast et al. (\cite{feastpl}) sources}
\label{feasttab}
\[
\begin{array}{lccrrrrrrrr}
\hline
\noalign{\smallskip}
\mathrm{Source} & \mathrm{Ra} & \mathrm{Dec} & \mathrm{Tile} & \mathrm{Seq.} & \mathrm{No.} & \mathrm{Dis.} & m_{BOL} & \mathrm{P} & \mathrm{A} & \mathrm{q} \\ 
 & \multicolumn{2}{c}{\mathrm{J2000}} & & & & & & \multicolumn{3}{c}{\mathrm{RED}} \\
\mathrm{W132}  & 5:28:54.13 & -69:40:15.33 & 77 & 7914 &   16 & 2.71 & 14.48 & 158 & 219 &  594 \\
\mathrm{W151}  & 5:29:28.84 & -69:32:53.79 & 77 & 7916 &   24 & 1.74 & 14.67 & 173 & 292 &  667 \\
\mathrm{W148}  & 5:29:25.19 & -69:26:54.50 & 77 & 7917 &   61 & 2.73 & 14.74 & 183 & 417 &  216 \\
\mathrm{W19}   & 5:26:23.89 & -69:34:00.44 & 77 & 7431 &   18 & 4.35 & 14.33 & 196 & 119 &  172 \\
\mathrm{W77}   & 5:28:6.25  & -69:32:28.82 & 77 & 7795 &   15 & 2.33 & 14.20 & 209 & 166 &  656 \\
\mathrm{W74}   & 5:27:51.13 & -69:58:09.88 & 77 & 7667 &  943 & 3.16 & 14.53 & 233 & 392 &  791 \\
\mathrm{W1}    & 5:25:49.46 & -69:44:36.01 & 77 & 7429 &  303 & 3.37 & 14.55 & 237 & 508 & 1142 \\
\mathrm{W140}  & 5:29:16.66 & -69:43:34.97 & 77 & 7913 &  369 & 1.93 & 14.17 & 243 & 365 & 1155 \\
\mathrm{W48}   & 5:27:10.38 & -69:28:27.76 & 77 & 7554 &   11 & 2.94 & 13.97 & 280 & 453 & 1251 \\
\mathrm{W46}   & 5:27:09.34 & -69:41:57.76 & 77 & 7550 &   22 & 2.50 & 14.20 & 297 & 107 &  274 \\
\mathrm{W126}  & 5:28:41.48 & -69:51:15.48 & 77 & 7790 &  310 & 3.22 & 13.91 & 315 & 612 & 1114 \\
\mathrm{W103}  & 5:28:27.36 & -69:49:03.44 & 77 & 7791 &  104 & 2.65 & 14.14 & 366 & 368 &  419 \\
\mathrm{GR5}   & 5:12:38.39 & -65:56:10.10 & 57 & 5308 &   13 & 2.85 & 15.05 & 140 & 394 &  176 \\
\mathrm{RGC60} & 5:26:34.72 & -67:51:58.68 &  4 & 7578 &   16 & 1.23 & 14.64 & 157 & 261 &  141 \\
\mathrm{GR27}  & 5:29:01.37 & -65:29:34.64 & 65 & 7977 &   11 & 0.63 & 14.31 & 194 & 241 &  108 \\
\mathrm{GR7}   & 5:07:08.26 & -66:35:23.75 & 56 & 4330 &  282 & 1.97 & 14.51 & 208 & 484 &  190 \\
\mathrm{RGC55} & 5:29:12.42 & -67:57:30.10 &  4 & 7940 &   12 & 2.42 & 14.25 & 283 & 170 &  156 \\
\mathrm{GR10}  & 5:05:02.81 & -66:53:10.92 & 53 & 4084 &   14 & 0.88 & 13.73 & 306 & 323 &  295 \\
\mathrm{GR26}  & 5:34:15.15 & -65:29:49.35 & 65 & 8823 &   24 & 1.01 & 14.28 & 306 & 248 &  131 \\
\mathrm{GR30}  & 5:24:40.13 & -65:41:18.83 & 63 & 7248 &   11 & 1.30 & 13.71 & 315 & 419 &  264 \\
\mathrm{W220}  & 5:30:43.00 & -70:02:39.33 & 77 & 8150 &   42 & 5.61 & 14.12 & 283 & 177 & 1121 \\ 
\mathrm{GR11}  & 5:02:52.87 & -67:07:40.78 & 25 & 3717 & 1133 & 0.06 & 13.92 & 322 & 184 &  282 \\
\mathrm{GR12}  & 5:03:49.17 & -66:15:57.55 & 55 & 3851 &   16 & 1.52 & 12.36 & 592 & 463 &  329 \\
\mathrm{GR1 }  & 5:15:40.39 & -66:04:59.15 & 59 & 5790 &   13 & 2.74 & 12.24 & 535 & 543 &  260 \\
\mathrm{GR48}  & 5:23:53.00 & -66:41:29.69 & 60 & 7112 &   15 & 1.67 & 11.87 & 634 & 505 &  217 \\
\mathrm{GR17}  & 5:39:32.97 & -66:56:35.78 & 67 & 9649 &   46 & 2.24 & 11.87 & 729 & 532 &  384 \\ 
\noalign{\smallskip}
\hline
\end{array}
\]
\end{table*}

\section{Bolometric correction(s)}

\subsection{BC$_I$ versus $I-J$}
\begin{figure}
\resizebox{\hsize}{!}{\includegraphics{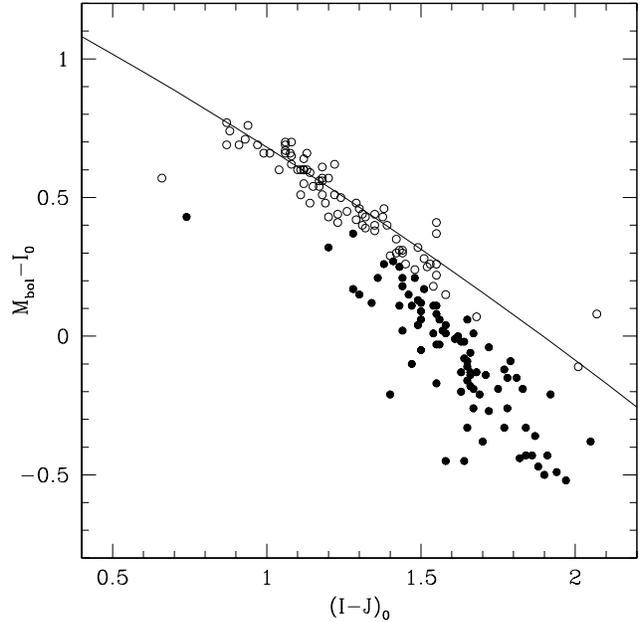}}
\caption{The bolometric correction BC$_I$ as a function of the $I-J$ colour for the O--rich (open circles) and C--rich (filled circles) variable stars in the sample. The continuous line is the relation derived by Alvarez et al. (\cite{alva}) for cool O--rich stars.}
\label{bci}
\end{figure}
Fig. \ref{bci}  compares the  bolometric correction obtained  from the
integration  under the  SED (Sect.  3.1) and  the relation  derived by
Alvarez et al. (\cite{alva}) for  cool O--rich stars. The agreement is
very good for the O--rich stars  of our sample. This confirms that the
relation  is  valid  for  the  SMC   as  well as  for  the  LMC  and  the
Bulge.  However  this  relation  overestimates  the  correction  for
C--rich variables in  the SMC. On the other hand  we should check that
the integration under the SED  does not underestimates the bolometric
correction (if the stars have a circumstellar envelope and dust extinction 
the correction for energy re--radiated by dust becomes important). 
  Our measurements have been corrected for interstellar extinction while
the  relation has not.  The agreement  for O--rich stars 
might be  a consequence  of the
systematic difference  between the DENIS  filters, used in  this work,
and  the broad  pass--bands  by  Bessels \&  Brett (\cite{besbre})  and
Bessel (\cite{besse}) used to derive the relation.
The disagreement for C--rich stars might be a consequence of the SiC 
dust feature which causes the spline not to follow the continuum of the 
SED. However the data of Alvarez et al. (\cite{alva}) include 
only $6$ C--rich stars and a total of four stars belonging to the 
LMC or the SMC. In addition the spectral range covered by their 
measurements is limited to $2.5 \mu$m. The zero--point of the 
bolometric scale is the same in both studies.

\subsection{BC$_K$ versus $J-K$}
\begin{figure}
\resizebox{\hsize}{!}{\includegraphics{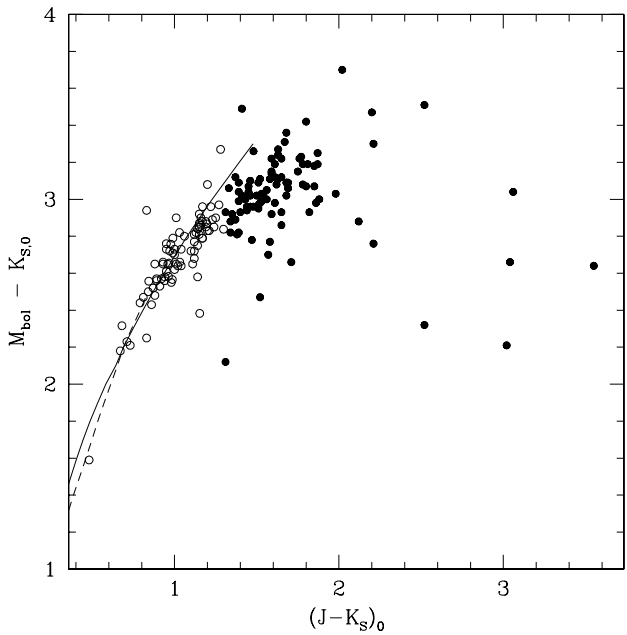}}
\caption{The bolometric correction BC$_K$ as a function of the $J-K_S$ colour for the O--rich (empty circles) and C--rich (filled circles) variable stars in the sample. The lines are the relations derived by Montegriffo et al. (\cite{monte}) for Population II giants of low (dashed line) and high metallicity (continuous line).}
\label{bck}
\end{figure}
Fig.  \ref{bck}  compares the  bolometric  correction  obtained as  in
Sect.   3.1   with   the    relation   derived   by   Montegriffo   et
al. (\cite{monte}) for Population II giants of low ([Fe/H]$<-1.0$) and
''high'' metallicity ([Fe/H]$>-1.0$). They used $M_{BOL,\sun}=4.75$. 
Their relation was 
defined up to $(J-K)=0.8$ but being the metal--poor relation very close to 
the metal--rich relation at redder colours we consider the latter as a  
possible extrapolation. This results in a fairly good fit for the O--rich 
variables while C--rich variables lie systematically below the relation. 
This is a similar effect as that observed for $BC_I$ versus $I-J$ 
(Fig. \ref{bci}). The giants in the sample of Montegriffo et al. 
(\cite{monte}) are members of galactic globular clusters. C--rich stars are 
rare members of these systems. The spectral range covered by the observations 
is limited to $3.5 \mu$m ($L$ band) and they derive the bolometric magnitude 
by integrating the SED which is previously re--constructed by means of a set 
of Plank functions at a given wavelength. These effects may influence the 
correction to be applied to the flux measured only in the $K$ band. 
Besides the $2.2 \mu$m filters used in both stu\-dies are not strictly the same.
We conclude that we cannot at present compare the correction for C--rich stars 
while O--rich stars are qualitatively in agreement. It would be useful to 
compare stars of other galaxies in the Local Group at a different metallicity.

\end{document}